# Dominance of Particle Resonances over Parametric Instabilities in High-Intensity Linear Accelerators

Dong-O Jeon[†] and Ji-Ho Jang

IRIS, Institute for Basic Science, Daejeon 34000, Republic of Korea

*ABSTRACT*

For high-intensity linear accelerators, space-charge halo mechanisms are largely classified into two families: particle resonances and parametric instabilities. The dominance between the fourth-order particle resonance and the envelope instability has been argued and studied. Our studies and previous literatures indicate the dominance of particle resonances over parametric instabilities in high-intensity linear accelerators. Any counter evidence has not been found yet. Also all the studies indicate that parametric instabilities are unlikely to be observed in actual linear accelerators unless waterbag or KV beams are generated. We propose a way to overcome the previous design rule to avoid the zero-current phase advance $\sigma_o > 90°$ for the high-intensity linac. The interplay is presented of the envelope instability and the fourth-order parametric instability.

PACS: 29.27.Bd, 41.75.-i

Keywords: high intensity beams, particle resonances, parametric instabilities;

# 1. Introduction

As the beam intensity increases in modern accelerators, self-field effects of the beam become significant and a lot of efforts have been taken to further their understanding. In 1959, Kapchinskij and Vladmirskij discovered a self-consistent beam distribution called "KV distribution" and derived the envelope equation [1]. In 1979, Haber et al. found a fourth order instability of KV distribution [2]. In 1983, Hofmann et al. derived various instabilities of the modes of KV distribution from Vlasov-Poisson equations and showed that $\sigma_o > 90°$ should be avoided in high-intensity linacs [3]. Here $\sigma_o$ is the zero current phase advance per cell. In 1984, Struckmeier et al. verified that $\sigma_o > 90°$ should be avoided in high-intensity linacs by solving the envelope equation for a mismatched beam [4]. These instabilities are recently called parametric instabilities, parametric resonances, structure resonances, or coherent resonances and so on, and the second order instability is widely known as the envelope instability. In order to avoid a confusion with "particle" parametric resonances such as in Refs. [5-8], these instabilities would better be called parametric instabilities rather than parametric resonances.

In 2009, Jeon et al. discovered the $4\sigma = 360°$ fourth-order particle resonance in high-intensity linear accelerators [9] and this linac resonance was experimentally verified in two experiments [10,11]. In 2015, Jeon et al. discovered the $6\sigma = 720°$ sixth-order particle resonance in high-intensity linear accelerators [12]. $\sigma$ represents depressed phase advance per cell. These fourth and sixth-order particle resonances are described by the Hamiltonian with the analytical expression for the space-charge potential of 2D Gaussian distribution [12]. For instance the fourth-order particle resonance is represented by Eq. (1,2):

$$H = \frac{p_x^2}{2} + K_x \frac{x^2}{2} - V_2 x^2 + V_4 x^4 - V_6 x^6 + \cdots \tag{1}$$

$$H = \nu_x J_x - \xi_x J_x + \alpha_{xx} \frac{J_x^2}{2} + G_4 J_x^2 \cos(4\varphi_x - \theta + \eta_4) + \cdots \tag{2}$$

where $\xi_x = \frac{1}{2\pi}\int_0^L V_2\beta_x ds$, $\alpha_{xx} = \frac{3}{2\pi}\int_0^L V_4\beta_x^2 ds$, $G_4 e^{i\eta_4} = \frac{1}{4\pi}\int_0^L V_4\beta_x^2 e^{i[4\chi_x-(4\nu_x-1)\theta]}ds$. In 2016, the eighth-order and tenth-order particle resonances were found by Hofmann [13]. These particle resonances are excited by the nonlinear multipoles of the space charge potential of non-KV distribution. Also particle-core model analyses with 2D Gaussian distribution self-field have been reported such as Refs. [22-24].

Tiefenback's thesis in 1986 reported an experiment observation of brief appearance of "four-pointed structure" (10-mA beam) in the σ_o = 98° and σ_o = 102° linear transport lattices, but misidentified it as "fourth-order structure resonance" referring to Refs. [2,3] (fourth-order parametric instability in this paper's terminology) [14]. On the other hand, Ref. [9] identified the fourth-order particle resonance in linear accelerators for the first time, which is the discovery. The "four-pointed structure" observed in Ref. [14] was not the fourth-order parametric instability, but the fourth-order particle resonance. Their difference is well illustrated in Fig. 5 for instance.

One needs to distinguish these two families of space charge halo mechanisms: 1) particle resonances (or resonances, incoherent resonances); 2) parametric instabilities (or parametric resonances, coherent resonances, structure resonances, etc.) of the beam eigenmodes. Clear distinction between these two families was presented in Ref. [15]: particle resonances have stable fixed points and FFT spectrum shows a clear resonance frequency component, while parametric instabilities do not have stable fixed points, nor resonance frequency component. These two families have their own framework of theories: Poisson-Vlasov equations for parametric instabilities; Hamiltonian formalism for particle resonances. So far there is no theoretical framework showing the interplay between particle resonances and parametric instabilities.

In the experiments, only the 4σ = 360° fourth-order particle resonance has been observed, not the envelope instability [10,11]. This opened up a question whether the fourth-order particle resonance dominates over the well-known envelope instability. In 2015 it was shown that the envelope instability emerges following the fourth-order particle resonance in a constant-$\sigma_o$ lattice for Gaussian distribution [16]. Soon afterward, it was demonstrated that the envelope instability is suppressed in a constant-σ lattice where the fourth-order particle resonance is maintained [17,18]. More details are in Sec. 2.

Even though parametric instabilities have been known for long, the cases of their manifestation are very limited for non-KV beam distribution. In Sec. 2, multiparticle simulation results along with previous literatures are presented showing that particle resonances dominate over parametric instabilities for non-KV distribution in high-intensity linear accelerators. Some may argue that this is too general a statement. However, no counter examples have been found yet. It is also discussed whether parametric instabilities can be observed in actual linear accelerators. In Sec. 3, ways to operate high intensity linacs in $\sigma_o > 90°$ are presented. In Sec. 4, the interplay between the envelope instability and the fourth-order parametric instability is presented for KV distribution. A case is reported where the envelope instability emerges without the preceding fourth-order parametric instability. In Sec. 5, mitigation ways of the envelope instability or the fourth-order particle resonance are presented. And the conclusion follows.

## 2. Particle resonances dominate over parametric instabilities

The dominance between the fourth-order particle resonance and the envelope instability has been a subject of several studies since the discovery of the fourth-order particle resonance [9,10,16-18]. Here, we take a further step and look into the dominance in general among particle resonances and parametric instabilities by putting together simulation results and previous literatures. The conclusion has never been claimed before.

It is informative to visualize the domains of particle resonances and of parametric instabilities observed in multi-particle simulations, as displayed in Fig. 1. The cases of parametric instabilities reported so far are very limited for non-KV distributions. The third and fourth-order parametric instabilities have been observed only for waterbag distribution away from the particle resonances [15-19], but not observed for Gaussian distribution [15]. Gaussian distribution has larger tune spread than waterbag distribution, and actual beam distribution in general has even larger tune spread because of more tails. Table I summarizes the parametric instabilities observed in multi-particle simulations vs. types of beam distribution. This suggests that these parametric instabilities are unlikely to be observed in actual linear accelerators. Besides the early experiment work also stated: our work shows no evidence of any unstable behavior for the lowest accessible values of $\sigma$ for lattice strength up to about $\sigma_o = 88°$ [14].

No other parametric instabilities except for the envelope instability have been manifested in the stopband of particle resonances for non-KV distribution. However, the envelope instability is manifested when the fourth-order particle resonance fades away as $\sigma$ increases and reaches 90° (primarily in a constant-$\sigma_o$ lattice). Even the envelope instability is suppressed in a constant-$\sigma$ lattice [17,18] where the fourth-order particle resonance is maintained.

TABLE I. Parametric instabilities and beam distribution in linear accelerators.

| Parametric instability | KV distribution | waterbag distribution | Gaussian distribution |
| --- | --- | --- | --- |
| envelope instability | O | O | O |
| 3$^{rd}$ order, 4$^{th}$ order | O | O | X |
| higher order | O | X | X |

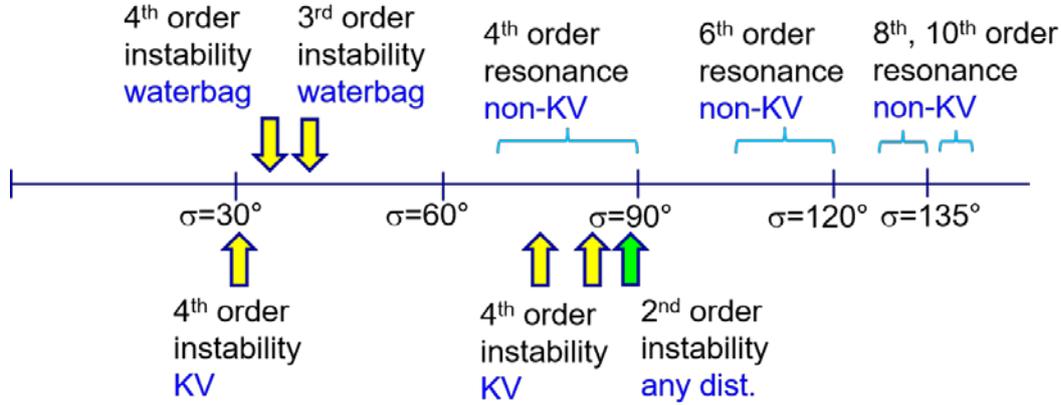

FIG. 1. Domain plot of particle resonances and parametric instabilities on the axis of depressed phase advance σ. Brackets represent the stopbands of particle resonances (their width is dependent on the beam current). Arrows represent the observed parametric instabilities in multiparticle simulations along with the distribution type. Parametric instabilities are observed only for waterbag distribution away from the particle resonances, and for KV distribution.

Numerical simulations are performed with initially well-matched Gaussian or KV distribution with 20 000 to 100 000 macro-particles using the PARMILA code with a 3D PIC space-charge routine [20]. 5 ~ 50 mA $^{40}Ar^{10+}$ beams with initial normalized rms emittances $\varepsilon_x = \varepsilon_y = 0.115$ [mm mrad], $\varepsilon_z = 0.120$ [mm mrad] are used for simulations. The beam has the initial energy of 5 MeV/u and is accelerated. A linac with an FOFODODO transverse focusing lattice is used for simulations. The acceleration voltage is applied in the drifts that are marked as "O" of the FOFODODO.

Just below phase advance σ = 90°, there are three mechanisms: the $4\sigma = 360°$ fourth-order particle resonance; the envelope instability; the $4(\sigma_o - \Delta\sigma_{4,coh}) = 360°$ fourth-order parametric instability. The dominance among these mechanisms, conditions for their manifestation and suppression are presented.

Firstly, we look into the dominance between the fourth-order particle resonance and the envelope instability in linacs using well-matched input Gaussian distribution. The $4\sigma = 360°$ fourth-order particle resonance was predicted in simulations with realistic beam distributions [9], and as predicted, only the fourth-order particle resonance was observed in the experiment, not the envelope instability [10]. Constant-$\sigma_o$ lattices were used in the experiment. Refs. [9,10] discussed the question whether the fourth-order particle resonance dominates over the envelope instability. Also the experiment and simulation study using the SNS linac showed the $4\sigma = 360°$ fourth-order particle resonance not the envelope instability when constant-$\sigma$ lattices were employed [11]. The SNS CCL (Coupled Cavity Linac) consists of 24 periods of FODO lattice. It should be noted that the studies in Refs. [9,11] used realistic beam distributions based on emittance measurements.

Ref. [16] reported that the envelope instability was manifested following the fourth-order particle resonance in a constant-$\sigma_o$ lattice for Gaussian distribution. Soon afterward, it was demonstrated that the envelope instability was manifested because the fourth-order particle resonance faded away (as $\sigma$ increases and reaches 90° in a constant-$\sigma_o$ lattice) [17,18]. Moreover, the fourth-order particle resonance dominates over the envelope instability in a constant-$\sigma$ lattice. As long as the fourth-order resonance is maintained by keeping $\sigma$ constant, the fourth-order resonance dominates and the envelope instability is suppressed, as shown in Figs. 2, 3.

Figure 2 displays the emittance growth factor ($\varepsilon_{final} / \varepsilon_{initial}$) over a wide range of phase advance and tune depression: phase advance $70° < \sigma < 95°$; initial tune depression $13° < (\sigma_o - \sigma) < 45°$ (corresponding to 5 ~ 40 mA $^{40}Ar^{10+}$ beams). Constant-$\sigma$ linac lattices are used and the fourth-order particle resonance dominates over the envelope instability at all the data points. None of the data shows a sign of the envelope instability. Emittance increase and beam distribution of one point in Fig. 2 (the σ = 87° lattice and initial tune depression 45° (40-mA case)) are portrayed in Fig. 3. The top plot shows the beam distribution and the emittance growth along the constant-σ lattice of 80 cells (or periods). Only the fourth-order particle resonance is manifested and the envelope instability is suppressed. The resultant rms emittance growth factor ($\varepsilon_{final} / \varepsilon_{initial}$) is 11.6. The bottom plot displays the corresponding depressed phase advance σ and zero-current phase advance $\sigma_o$.

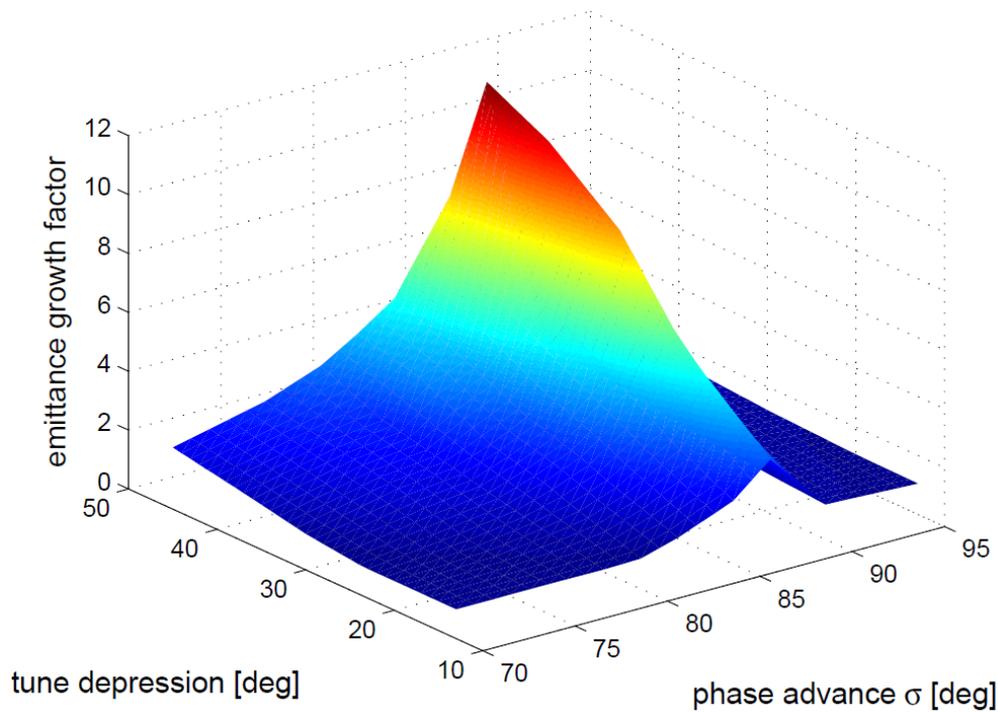

FIG. 2. 3D plot of emittance growth factor (= $\varepsilon_{final} / \varepsilon_{initial}$) vs. the initial tune depression ($\sigma_o - \sigma$) and phase advance σ of linac lattices. For each data point, a constant-σ lattice is used. The initial tune depression corresponds to 5 ~ 40 mA $^{40}Ar^{10+}$ beams. Only the fourth-order particle resonance is manifested and the envelope instability suppressed. And there is no resonance effect observed in σ > 90°. For the simulations, initially well-matched Gaussian distribution is used.

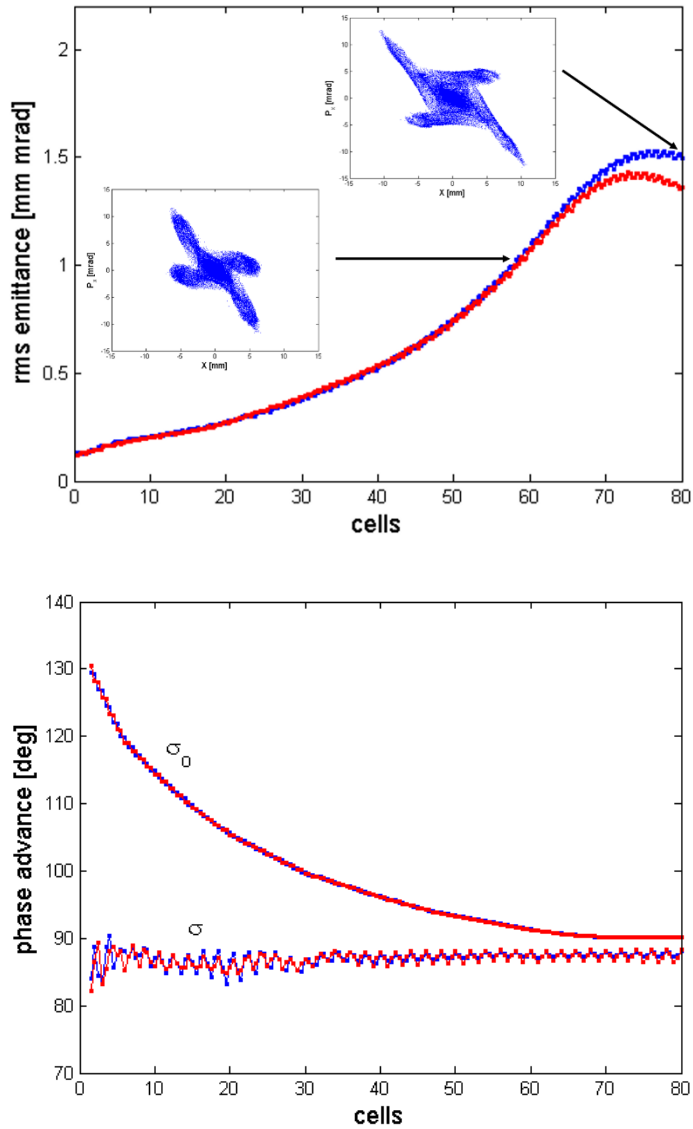

FIG. 3. Plots of normalized rms emittances (top plot); plots of $\sigma_o$ and $\sigma$ in transverse planes (bottom plot) over 80 cells. The linac lattice maintains $\sigma \approx 87°$ along the linac. As a result, the fourth-order particle resonance dominates and persists throughout the linac as shown in the top plot. There is no sign of the envelope instability. Blue (red) curves represent quantities in the X (Y) plane. Well-matched 40-mA Gaussian distribution is used, and the initial tune depression is $(\sigma_o - \sigma) \approx 45°$.

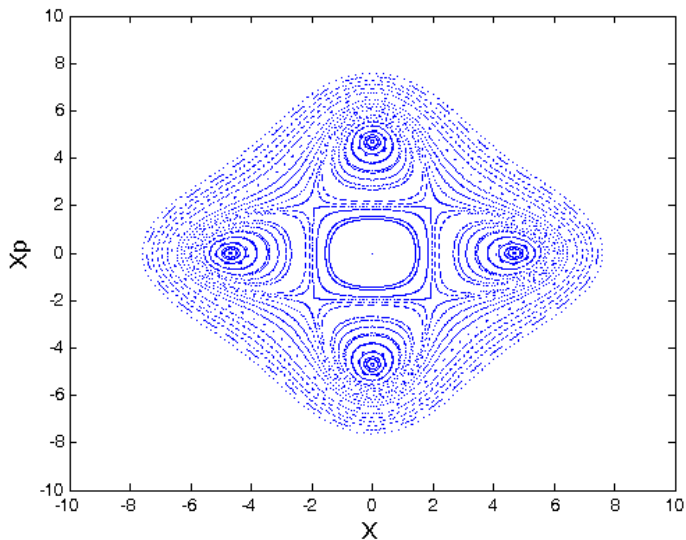

FIG. 4. Plot of typical resonance island structure of the fourth-order particle resonance. There are four stable islands around four stable fixed points, and stable islands represent potential wells that hold beam particles.

These demonstrate that the fourth-order particle resonance dominates over the envelope instability and is the limiting mechanism in $\sigma_o > 90°$. In Sec. 5, ways to avoid or to mitigate the envelope instability are summarized. It needs to be reminded that stable fixed points and resonance islands form as shown in Fig. 4, when the fourth-order particle resonance is induced. Resonance islands represent potential wells in the phase space that hold beam particles. Fixing $\sigma$ of a linac lattice preserves resonance islands in the phase space, and suppresses the excitation of the envelope instability.

Secondly, we look into the dominance between the $4\sigma = 360°$ fourth-order particle resonance and the $4(\sigma_o - \Delta\sigma_{4,coh}) = 360°$ fourth-order parametric instability. Even though the appearance of the particle resonance and that of the parametric instability look alike as displayed in Fig. 5, they are quite different mechanisms. The core difference between them is whether the FFT resonance frequency component exists as demonstrated in Ref. [15]. The fourth-order particle resonance always comes with the resonance frequency component in the FFT spectrum (due to stable fixed points). For any non-KV distribution (including realistic beam distribution), the fourth-order particle resonance rather than the fourth-order instabilities gets excited in the stopband ($90° < \sigma_o$ and $\sigma < 90°$): realistic distributions [9-11]; Gaussian and waterbag distributions [15-19, 22-23]. No counter example has been found yet. Thus it is apparent that for non-KV distribution, the fourth-order particle resonance dominates over the $4(\sigma_o - \Delta\sigma_{4,coh}) = 360°$ fourth-order parametric instability. The $4(\sigma_o - \Delta\sigma_{4,coh}) = 360°$ fourth-order parametric instability is manifested below $\sigma = 90°$ only for KV distribution [15,19].

Thirdly, the $6\sigma = 720°$ sixth-order particle resonance dominates over the $3(\sigma_o - \Delta\sigma_{3,coh}) = 360°$ third-order parametric instability. The $3(\sigma_o - \Delta\sigma_{3,coh}) = 360°$ third-order instability has not been observed for non-KV distributions such as waterbag or Gaussian distribution in the range just below $\sigma = 120°$. Instead, the $6\sigma = 720°$ sixth-order particle resonance is observed [12,19]. Ref. [12] employed constant-$\sigma$ lattices and ramping-$\sigma$ lattices. Also the third-order parametric instability is not observed in Fig. 7 where a constant-$\sigma_o$ lattice is used.

Fourthly, it should be noted that the cases of the third and fourth-order parametric instabilities for non-KV distribution are very limited, and that they are manifested away from the particle resonances. The $4(\sigma_o - \Delta\sigma_{4,coh}) = 180°$ fourth-order parametric instability has been observed for KV distribution (when $\sigma_o = 90°$ and $\sigma = 30°$) [2] and also for waterbag distribution (when $\sigma_o = 70°$ and $\sigma = 35°$) [19]. The $3(\sigma_o - \Delta\sigma_{3,coh}) = 180°$ third-order instability has also been observed for waterbag distribution (when $\sigma_o = 92°$ and $\sigma = 40°$) [16,17]. These parametric instabilities are observed away from $\sigma = 90°$ for waterbag distribution, but not for Gaussian distribution [15]. Parametric instabilities except for the envelope instability can be disregarded in actual linacs, unless KV or waterbag distribution is generated at the ion source. Near these instabilities, there might exist the $6\sigma = 360°$ sixth-order particle resonance (just below $\sigma = 60°$).

But this resonance has been reported to be very weak [12,19] unlike the 6σ = 720° sixth-order particle resonance, and does not seem to influence these parametric instabilities.

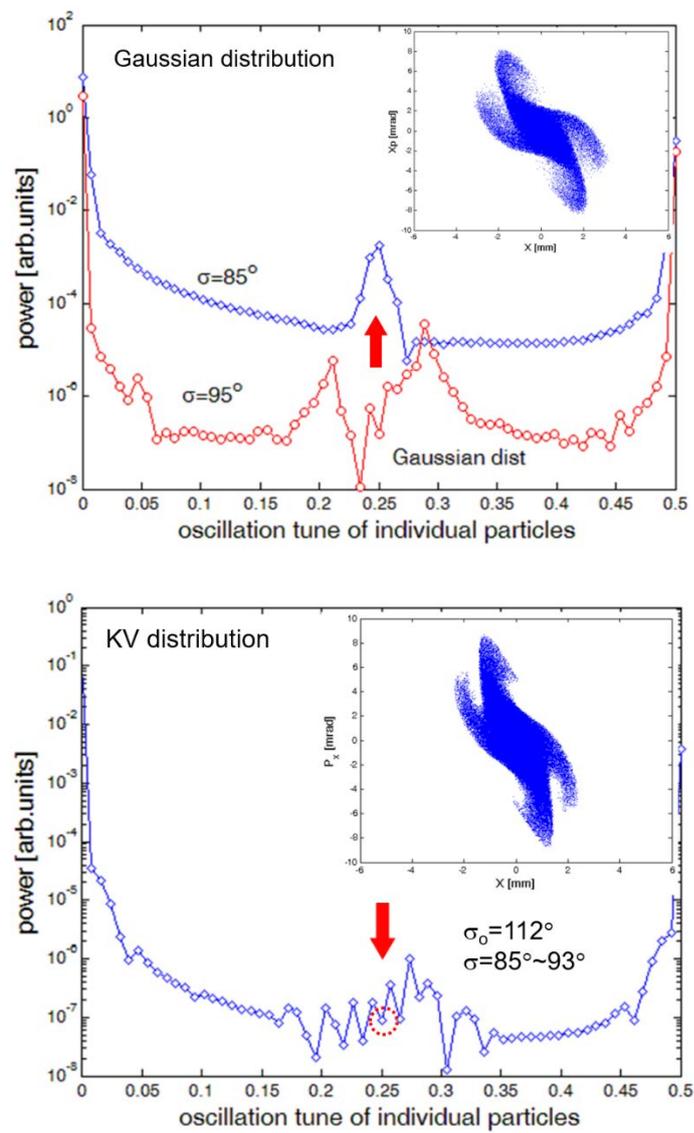

FIG. 5. FFT spectra and beam distribution of the fourth-order particle resonance (top plot); those of the fourth-order parametric instability (bottom plot). A clear resonance frequency component is observed at 1/4 for the σ = 85° case in the top plot. In case of the fourth-order parametric instability, no resonance frequency component is observed at 1/4. For the fourth-order parametric instability, a constant-$\sigma_o$ linac is used with $\sigma_o$ = 112°.

Lastly, it is how σ and $σ_o$ vary in the linac that determines the dynamics, independent of lattice types and beam distributions. Lattice type affects the degree of the beam flutter that determines the strength of the space charge effect. Lattice type in Ref. [9] was FOFODODO and that in Ref. [11] FODO. Refs. [22,23] used solenoid and quadrupole doublet lattices, and the fourth-order particle resonance and the envelope instability were observed according to the values of σ and $σ_o$. Most of studies used Gaussian, waterbag or KV distributions. And the studies in [9,11] are done with realistic beam distributions based on emittance measurements in the upstream linac.

## 3. Operating the high-intensity linac in σ > 90°

The region $σ_o$ > 90° has been regarded forbidden in high-intensity linac design and operation. As a possibility to explore the linac operation in $σ_o$ > 90°, it was proposed recently that introducing beam spinning can reduce the effects of the fourth-order particle resonance and the envelope instability [22,23]. Here the possibility is explored further to operate high-intensity linear accelerators in $σ_o$ > 90°.

We propose to operate the linac in σ > 90°, by which one can avoid the fourth-order particle resonance and the envelope instability. It should be noted that neither the fourth-order particle resonance [9] nor the envelope instability is manifested in σ > 90° (Fig. 3 of Ref. [19]). If one operates a constant-σ linac with σ slightly above 90° (such as 92°), then one can minimize the effect of the 6σ = 720° sixth-order resonance as well. The sixth-order resonance is relatively weak and about 17% emittance growth is reported for a constant σ = 112° linac with 10-mA $^{40}Ar^{10+}$ beam [12]. Figure 6 shows the rms emittance increase and the beam distribution in a constant σ = 92° linac with well-matched 50-mA Gaussian input distribution. The beam distribution does not show any sign of the fourth-order particle resonance, nor the envelope instability, nor the sixth-order particle resonance. The resultant rms emittance growth factor ($\varepsilon_{final}$ / $\varepsilon_{initial}$) is 2.01 which is small compared with the emittance growth of 11.6 in the case of constant $\sigma = 87°$ lattice, 40-mA $^{40}Ar^{10+}$ beam in Fig. 3. This emittance growth which increases for higher beam current is mainly attributed to the redistribution from Gaussian to the equilibrium distribution.

Also one can use a constant-$\sigma_o$ linac or a variation (while keeping $\sigma > 90°$) in order to pass through the $6\sigma = 720°$ sixth-order resonance rather quickly. As illustrated in Fig. 7, in a constant-$\sigma_o$ lattice, the sixth-order particle resonance does not induce significant emittance growth, because the resonance is relatively weak and because the beam passes through the resonance briefly ($\sigma$ increases from 106° to 119°). In a constant $\sigma_o = 140°$ linac, six-fold structures briefly appear around $\sigma = 110°$ and soon disappears as $\sigma$ increases to 120° (stable fixed points move toward the origin and the resonance disappears). Well-matched 20-mA Gaussian input distribution is used. In Fig. 6 and Fig. 7, neither the envelope instability nor the fourth-order particle resonance is manifested.

Operating a linac in $\sigma > 90°$ is feasible and has advantages: 1) it avoids the fourth-order resonance and the envelope instability; 2) does not require additional hardware (just optics change). And it can be useful when relatively small beam size is required in operation. Moreover, enough parameter space is available: $\sigma$ from 90° to 125° or higher. The upper limit needs to be explored.

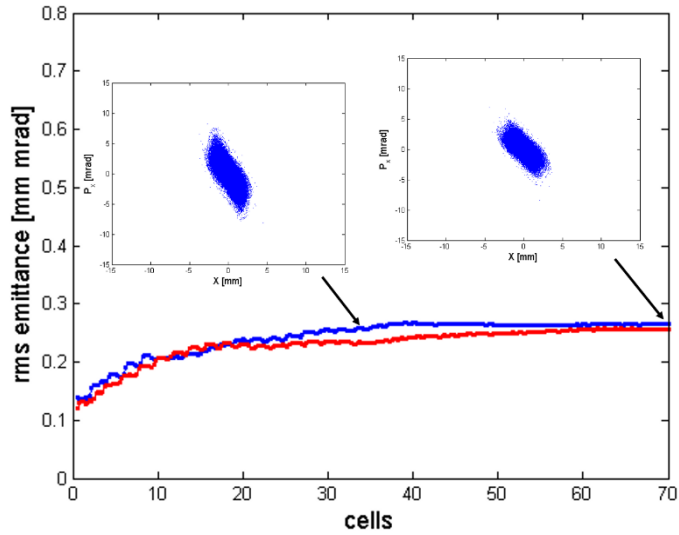

FIG. 6. Plots of transverse rms emittances and beam distribution in a constant $\sigma = 92°$ linac. Corresponding $\sigma_o$ decreases from 136° to 113°. Well-matched $^{40}Ar^{10+}$ 50-mA Gaussian input distribution is used. The beam distribution does not show a sign of the fourth order resonance nor the envelope instability. The rms emittance growth factor ($\varepsilon_{final} / \varepsilon_{initial}$) is 2.01 which is small compared with the emittance growth in a constant $\sigma = 87°$ linac, 40-mA beam in Fig. 3. This emittance growth is mainly attributed to the redistribution from Gaussian to the equilibrium distribution. Emittance growth is bigger than in Fig. 7 because of higher beam current. Blue (red) curves represent quantities in the X (Y) plane.

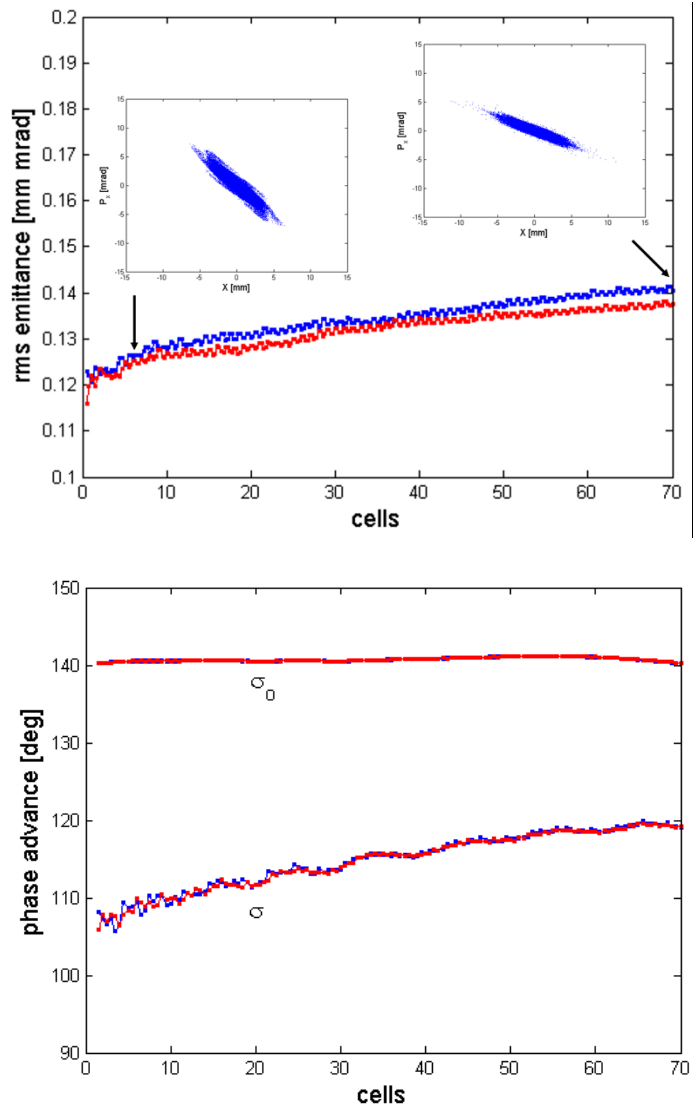

FIG. 7. Transverse rms emittances and beam distribution in a constant $\sigma_o = 140°$ linac (top plot); depressed and zero-current phase advances $\sigma$ and $\sigma_o$ (bottom plot). Well-matched $^{40}Ar^{10+}$ 20-mA Gaussian input distribution is used. Six-fold structure briefly appears around $\sigma = 110°$ but soon disappears as $\sigma$ increases to 120° (stable fixed points move toward the origin and the resonance fades away). About 16% emittance growth is induced and no sign of the envelope instability is observed in the beam distribution. Blue (red) curves represent quantities in the X (Y) plane.

## 4. The fourth-order instability and the envelope instability

The interplay is studied of the $4(\sigma_o - \Delta\sigma_{4,coh}) = 360°$ fourth-order parametric instability and the envelope instability for KV distribution. Figure 8 portrays the stopbands and emittance growth of the fourth-order parametric instability (blue curve) and of the fourth-order particle resonance (red curve) for 10-mA $^{40}Ar^{10+}$ KV and Gaussian distribution respectively. Horizontal axis represents the σ value of a lattice (constant-σ lattices are used). Emittance growth at the $28^{th}$ cell is plotted to exclude the envelope instability contribution.

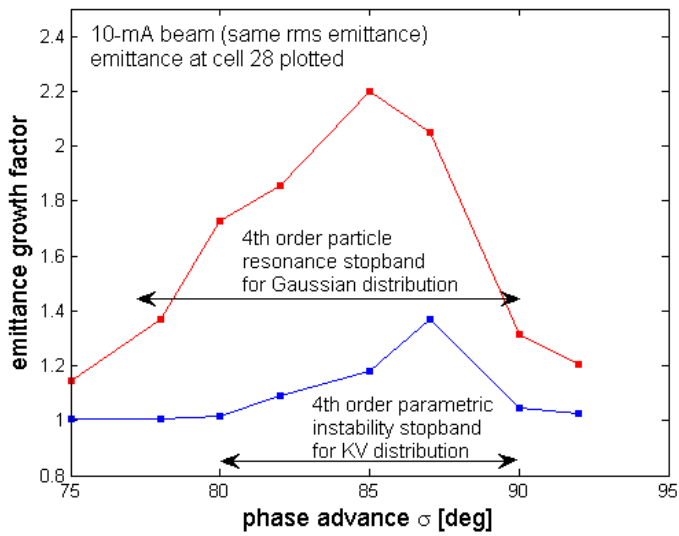

FIG. 8. Stopband and emittance growth of the fourth-order particle resonance and of the $4(\sigma_o - \Delta\sigma_{4,coh}) = 360°$ fourth-order parametric instability. Horizontal axis represents the σ value of a lattice (constant-σ lattices are used). It is observed that the stopband of the particle resonance is wider than that of the parametric resonance. For the particle resonance, $^{40}Ar^{10+}$ 10-mA Gaussian distribution is used; for the parametric instability, $^{40}Ar^{10+}$ 10-mA KV distribution is used with the same rms emittances. Emittance growth at the $28^{th}$ cell is plotted to exclude the envelope instability contribution.

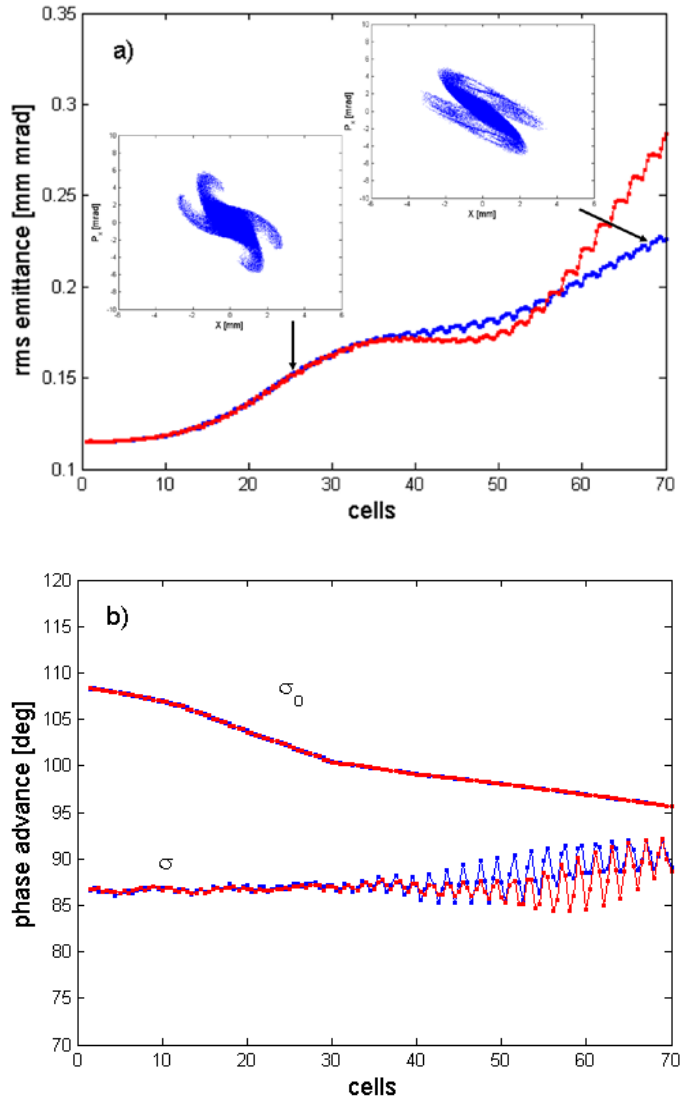

FIG. 9. Emittance growth and the beam distribution (top plot a); the corresponding phase advances $\sigma$ and $\sigma_o$ (bottom plot b) in a constant $\sigma = 87°$ lattice (a point of the blue curve in Fig. 8). A well-matched, $^{40}Ar^{10+}$ 10-mA input KV distribution is used. The $4(\sigma_o - \Delta\sigma_{4,coh}) = 360°$ fourth-order parametric instability emerges first and then the envelope instability follows. Blue (red) curves represent quantities in the X (Y) plane.

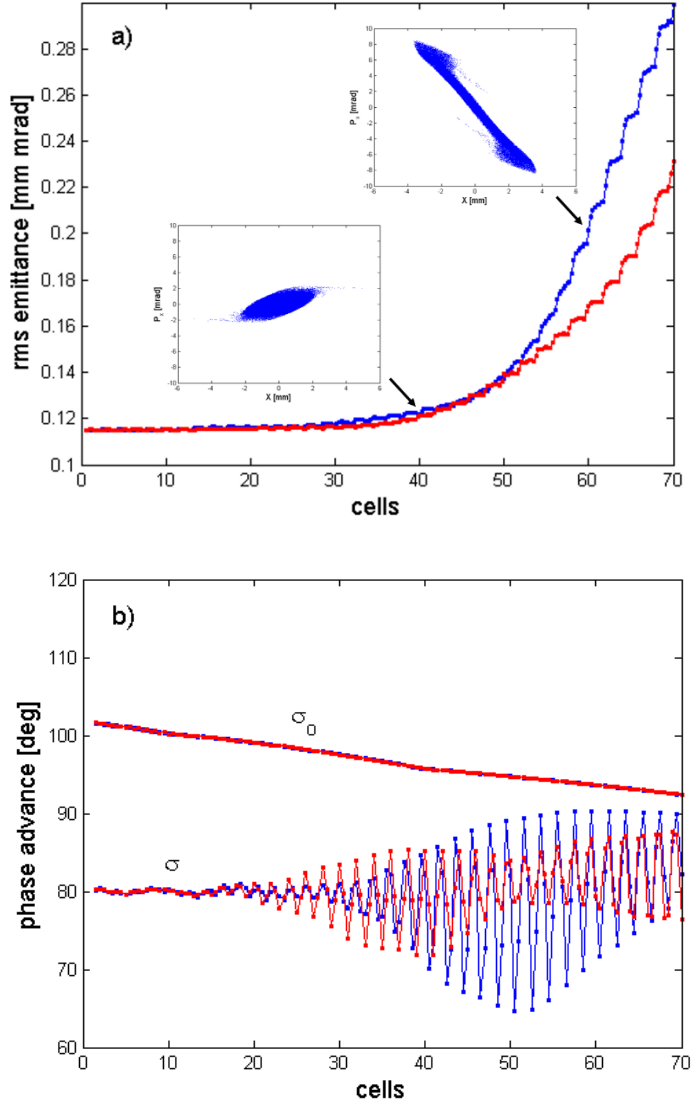

FIG. 10. Emittance growth and the beam distribution (top plot a); the corresponding phase advances $\sigma$ and $\sigma_o$ (bottom plot b) in a constant $\sigma = 80°$ lattice (a point of the blue curve in Fig. 8). Well-matched, $^{40}Ar^{10+}$ 10-mA input KV distribution is used. Because this operating point is just outside the instability stopband, the $4(\sigma_o - \Delta\sigma_{4,coh}) = 360°$ fourth-order parametric instability does not show up and the envelope instability alone is manifested. Blue (red) curves represent quantities in the X (Y) plane.

As depicted in Fig. 9, for initially well-matched KV distribution, the fourth-order parametric instability appears first and the envelope instability follows inside the fourth-order instability stopband. However, at σ = 80° (the lattice with σ = 80°) which lies just outside of the fourth-order instability stopband, the envelope instability emerges without the fourth-order instability preceding, as shown in Fig. 10. In this case the envelope instability emerges from the noise of KV beam distribution. For the fourth-order instability with KV distribution, the envelope instability is manifested even in constant-σ lattices. Using a constant-σ lattice does not suppress the envelope instability because there are no particle resonance islands to the fourth-order parametric instability unlike the fourth-order particle resonance.

## 5. Mitigation of the envelope instability or the fourth-order resonance

In the past, the envelope instability was regarded as the most dangerous mechanism and the linac designs have avoided the region $\sigma_o > 90°$ [3,4]. However, recent studies tend to point that the envelope instability can be suppressed or mitigated in various ways. The proposal of running the linac with σ > 90° is to avoid the fourth-order particle resonance and the envelope instability, as discussed in Section 3. Figure 6 shows the rms emittance increase and the beam distribution in a constant σ = 92° lattice, and Fig. 7 displays results in a constant $\sigma_o = 140°$ lattice while keeping σ > 90°. Running the linac with σ > 90° turns out feasible and induces only minimal emittance increase.

Ways to suppress or mitigate the envelope instability that have been reported earlier are listed below. As shown in Refs. [17,18] and in Sec. 2, the envelope instability can be suppressed by adopting a constant-σ linac lattice when σ < 90°. The envelope instability and associated emittance growth can be suppressed with a constant-σ lattice in the region of the fourth-order resonance and the envelope instability. In the fourth-order particle resonance experiment of Ref. [11], a constant-σ lattice was already applied in the CCL (Coupled Cavity Linac) of the SNS linac in order to maintain a constant resonance condition. In the experiment and the related simulations, no sign of the envelope instability was observed.

It was also reported by Qiang that fast acceleration mitigates the envelope instability [21]. As portrayed in Fig. 11, fast acceleration pushes the beam through the stopband quickly and reduces the emittance growth in a constant-$\sigma_o$ lattice [23].

Recent studies explored the possibility of mitigating the fourth-order particle resonance and the envelope instability by employing the beam spinning [22,23]. It was shown that the emittance growth by the fourth-order particle resonance and the envelope instability is reduced considerably, through enhancing the X-Y coupling in the space-charge potential by beam spinning.

These results show that the envelope instability can be suppressed or mitigated. In this light, the envelope instability is not an essential hurdle and the fourth-order resonance is the limiting mechanism when operating the high-intensity linac in $\sigma_o > 90°$.

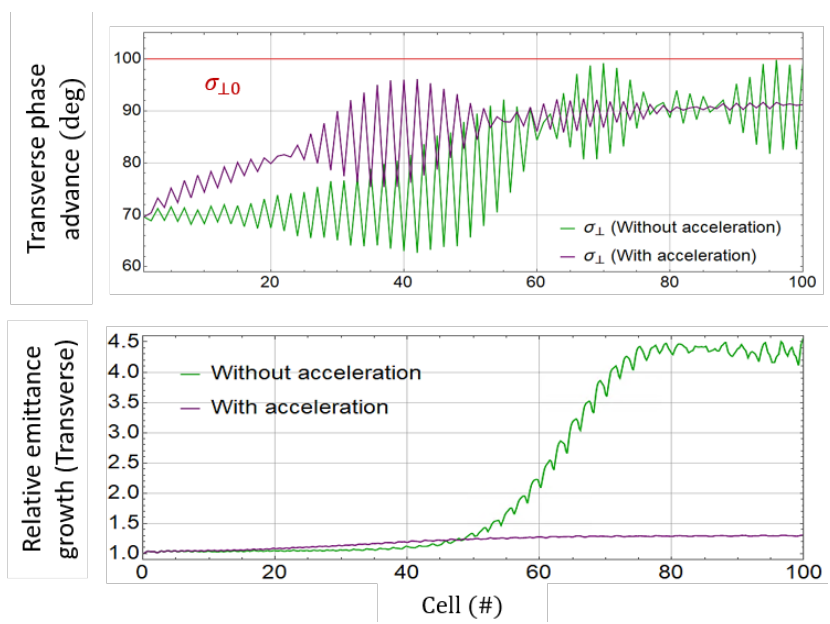

FIG. 11. Transverse phase advances (top plot) and the relative rms emittance growth (bottom plot) for two cases: with and without acceleration. With acceleration, the emittance growth is reduced significantly. Constant-$\sigma_o$ lattice is used. (Courtesy of [Y.L. Cheon et al., Phys. Rev. Accel. Beams **25**, 064002 (2022)]).

## 6. Conclusion

Studies and previous literatures indicate that particle resonances dominate over parametric instabilities in high-intensity linear accelerators in the stopbands of particle resonances for non-KV distribution. Parametric instabilities except for the envelope instability are unlikely to be observed in actual linacs, unless KV or waterbag distribution is generated at the ion source. Even the envelope instability can be suppressed when a constant-$\sigma$ lattice is employed. The envelope instability is manifested following the fourth-order particle resonance simply because the fourth-order resonance fades away, as $\sigma$ increases and reaches 90° in a constant-$\sigma_o$ lattice.

Operating the high-intensity linac in $\sigma > 90°$ avoids the fourth-order particle resonance and the envelope instability, overcoming the previous design rule to avoid the zero-current phase advance $\sigma_o > 90°$. The envelope instability can be suppressed or mitigated in a few ways: running the linac with $\sigma > 90°$; running the constant-$\sigma$ linac when $\sigma < 90°$; fast acceleration; beam spinning.

The stopband of the fourth-order particle resonance (for Gaussian distribution) is wider than that of the fourth-order parametric instability (for KV distribution) for the same beam current and same rms emittances. The $4(\sigma_o - \Delta\sigma_{4,coh}) = 360°$ fourth-order parametric instability precedes the envelope instability inside the fourth-order instability stopband, whereas only the envelope instability is manifested outside the stopband.

## Acknowledgements

This work was supported by the Institute for Basic Science (IBS-I001-D1).